\documentstyle[12pt,aaspp4]{article}

\begin{document}

\begin {flushright}
November 21, 2000\\
Submitted to ApJ
\end{flushright}

\title{Stellar Kinematics of the Spiral Structure in the Solar Neighborhood}
 
\author{Taihei Yano\altaffilmark{1},
\altaffiltext{1}{Research Fellow of the Japan 
Society for the Promotion of Science.} Masashi Chiba and Naoteru Gouda}  

\affil{National Astronomical Observatory, 
Mitaka, Tokyo 181-8588, Japan\\
E-mail: yano@pluto.mtk.nao.ac.jp, chibams@gala.mtk.nao.ac.jp, naoteru.gouda@nao.ac.jp}

\begin{abstract}
We present a method, based on the kinematic analysis of the Galactic disk
stars, to clarify whether the internal motions of the stellar system in spiral
arms follow those expected in the density wave theory. The method relies on
the comparison with the linear relation between the phases of spatial
positions and epicyclic motions of stars, as drawn from the theory.
The application of the method to the 78 Galactic Cepheids near the Sun,
for which accurate proper motions are available from the {\it Hipparcos}
Catalogue, has revealed that these Cepheids hold no correlation between both
phases, thereby implying that their motions are in contradiction with the
theoretical predictions. Possible reasons for this discrepancy are discussed
and future prospects are outlined.
\end{abstract}

{\it Subject headings}:  
Galaxy: kinematics - spiral arm - density wave - Hipparcos

\section{Introduction}
Spiral structures of galaxies have been studied for a long time in order to
understand how these structures are formed (e.g., Roberts, Roberts \& Shu
1975; Rohlfs 1977; Binney \& Tremaine 1987). One of the proposed models
to explain spiral arms is that they are just material arms, where the stars
originally making up a spiral arm remains in the arm even at the later time.
However this simple model holds a wellknown problem which is called ``winding
problem'': the differential rotation in galactic disks winds up the arm
in a short time compared with the age of galaxies, so that the spiral pattern
would be too tightly wound compared with the observed spiral structures.
In contrast, the currently most popular model, which is free from the winding
problem, is the density wave theory (Lin \& Shu 1964), where a spiral arm is
regarded as a wave and wavelike oscillation of stellar motions propagates
through galactic disks. In this picture, the global spiral pattern is sustained
independently of individual stars moving at different angular velocities.
For a comprehensive review of the density wave theory, see, e.g., Rohlfs (1977)
and Binney \& Tremaine (1987).

The density wave theory has been suggested by various
observational aspects in spiral galaxies, including the relative distributions
of dust lane, interstellar gas, and H II regions across the arms (Fujimoto 1968;
Roberts 1969; Rohlfs 1977), intensity distribution of radio continuum radiation
(Mathewson, van der Kruit, \& Brouw 1972), and systematic variation of gaseous
velocity fields near the arms (e.g., Visser 1980). In particular,
recent high-resolution observations using CO emission have revealed detailed
streaming motions of molecular gas, which are generally in agreement with
predictions of the density wave models (e.g., Kuno \& Nakai 1997; Aalto et al.
1999). However, we note that these observational results provide only an
outcome of nonlinear interaction between interstellar matter and background
stellar arms, and it is yet unknown whether the motions of stars {\it themselves},
which make up spiral pattern, actually follow those predicted by the density
wave theory. In this regard, the direct access to detailed stellar kinematics
in disks is possible only in our Galaxy.

Here, we present a method to clarify this issue, based on the analysis of
local kinematics of disk stars. We then apply the method to 78 Cepheids
in the solar neighborhood, for which the precise data of proper motions are
available from the {\it Hipparcos} Catalogue (ESA 1997). Also, the distances
to these sample stars can be accurately estimated from the period-luminosity
relation, so that combined with the radial velocity data, the full
three-dimensional velocities are available. We note here that although we
focus on the local kinematics of spiral arms in this work, the method we
develop here can be applied to the motions of more remote stars distributed
over a whole disk, for which precise astrometric data will be provided by
the next-generation satellites such as {\it FAME} and {\it GAIA}.

Our paper is organized as follows. In \S 2, we describe the method to determine
whether or not the motions of stars agree with those expected in the density
wave theory. In \S 3, we show the detail of the sample stars and the
fundamental parameters of our Galaxy adopted in this work.
The application of our method to the sample stars is shown in \S 4.
Finally, \S 5 is devoted to discussion and conclusions.

\section{METHOD}
All orbits of the disk stars in our Galaxy are not perfectly circular. 
The discrepancy between the motion of a star and its reference circular orbit,
called orbit of the guiding center, can be represented by an epicyclic
motion of the star (Binney \& Tremaine 1987).
We describe here a method to compare the epicyclic motions of the stars
observed near the Sun with those expected in the density wave theory.

First, we define the ``position phase'', $\chi$, as a function of the position
of a star in the Galactic plane, by assuming that the shape of the spiral arm
is logarithmic as follows:
\begin{equation}
\chi = \ln \frac{(r/r_g)}{\tan i} - \theta + \Omega_p t \ ,
\label{1}
\end{equation}
where $r$ is the distance between the star and the Galactic Center (GC),
$r_g$ is the distance between the Sun and GC, $i$ is the pitch angle of the
spiral arm, $\theta$ is the angle between $r$ and $r_g$,
and $\Omega_p$ is the angular velocity of the spiral pattern in the azimuthal
direction. We note that all the stars on the same arm have the same value of
$\chi$ (Figure 1).

Second, we define the phase of the epicyclic motion of a star, $\phi$,
which is the angle between the velocity vector of the epicyclic motion
and the direction perpendicular to a spiral arm (Figure 1).
Then, if the stellar motions in the arm obey the density wave theory,
all the stars on the same arm have the same value of $\phi$
as well as the same value of $\chi$. Also, when the number of arms is $m$,
the epicyclic frequency $\kappa$ of the stellar motions is equal to $m$ times
as large as the angular frequency $\Omega-\Omega_p$
[$\kappa = m(\Omega-\Omega_p)$]. If so, since $d\phi/dt = -\kappa$,
and $d\chi /dt = \Omega_p-d\theta /dt = \Omega_p-\Omega$ from equation (\ref{1}),
we obtain the following linear relation between $\phi$ and $\chi$,
\begin{equation}
\phi \equiv m\chi \pmod{2\pi},
\label{2}
\end{equation}
if the stellar motions obey the density wave theory.

Thus, by assessing this linear relation between observed $\phi$ and $\chi$,
it is possible to investigate whether or not the observed stellar motions
follow those predicted by the density way theory. We can also derive
the number of the arms, $m$, from the slope in the $\phi-\chi$ relation.
In Figure 3, we show the case $m=4$ (solid lines) as an example.

\section{DATA}
We adopt the Galactic Cepheids as the tracers of stellar motions in spiral
arms, because these bright stars can be seen from a long distance and so
we can investigate a wide region around the Sun. Also, these young populations
show only a small deviation from circular rotation, thereby allowing us to
analyze the internal kinematics of spiral arms alone; relatively old stars,
such as dwarf stars, have too large velocity dispersions, possibly due to
repeated gravitational interactions with massive clouds (Spitzer \&
Schwarzschild 1953) in addition
to the effect of spiral arms. Furthermore we can obtain accurate distances for
Cepheids, based on the relation between pulsation period and absolute
magnitude. 
We do not use the {\it Hipparcos} parallaxes, which have generally large
errors for many Cepheids beyond $\sim 100$ pc from the Sun.
We adopt the Cepheid catalog compiled by Mishurov et al. (1997) for distances
and radial velocities, and the {\it Hipparcos} Catalogue for proper motions,
to calculate the individual motions of the Cepheids in the Galactic plane.

In the Mishurov et al. catalog, Cepheids in the region of $r >4$ kpc and
those in a binary system are excluded. Also, nearby Cepheids in the region
of $r < 0.5$ kpc are excluded in order to reduce the local effects like
Gould's Belt. Furthermore, Cepheids whose pulsation periods exceed 9 days
are also excluded, because they are supposed to be extremely young objects.
In addition, we further exclude the Cepheids within 1 kpc for the following
reason. When we compare the observed distribution of Cepheids in the
$\phi-\chi$ plane with the theoretical prediction in a quantitative manner,
as will be described later, we assign larger statistical weights to the
Cepheids having a smaller observational error [eq. (\ref{3})]. This leads to
larger weights to the Cepheids located close to the Sun, say $r < 1$ kpc,
so that the result of the analysis will be largely determined by only small
number of Cepheids in a small region near the Sun. 
We also exclude two Cepheids whose peculiar velocity is exceptionally large,
over 50 km s$^{-1}$, compared to other ones.
As a consequence, we adopt 78 Cepheids in this work, and their spatial
distribution is shown in Figure 2.

We adopt $r_g=8.3$ kpc in our analysis, which is approximately an average
of observed values ranging from 8.1 kpc to 8.5 kpc (Kerr \& Lynden-Bell 1986;
Hanson 1987; Pont, Mayor \& Burk 1994; Feast \& Whitelock 1997).
As for the pitch angle of spiral arms, $i$, in our Galaxy, several authors
investigated one of the conspicuous spiral arms near the Sun, the Sagittarius
arm, based on the spatial distributions of open clusters, CO emissions, or
O-B2 clusters, and arrived at several to about $20^{\circ}$ (Pavlovskaya \&
Suchkov 1984; Dame et al. 1986; Grabelsky et al. 1988; Alfaro, Cabrera-Cano
\& Delgado 1992). In this work, we adopt $i = -8.0^{\circ}$ in the
solar neighborhood, where the negative value for $i$ denotes a trailing arm.
We have found that even if we change the values of these parameters over a
likely range, the result shown below remains essentially unchanged.

In order to analyze the epicyclic motion of the Cepheids, we require to
subtract both the local solar motion with respect to the local standard of rest
and the effect of the Galactic differential rotation from the observed motions.
As the local solar motion, we adopt 15.5 km s$^{-1}$ in the direction to
$l_{\sun}$=45$^{\circ}$ and $b_{\sun}=23.6^{\circ}$ (Kulikovskij 1985).
To estimate the effect of the Galactic differential rotation,
we assume that in the concerned region within about $4$ kpc from the Sun,
the Galactic rotation is monotonously changing with distance from the Sun,
where the Oort constant $A$ is assumed to be $13.0$ km s$^{-1}$kpc$^{-1}$.
We again note that the result shown in \S 4 is essentially independent of
the value of $A$ or the assumption for differential rotation.

\section{RESULTS}
We present the relation between the phases of the epicyclic motions of
the Cepheids $\phi$ and their position phases $\chi$ in Figure 3. We also show
the case for the 4-armed galaxy which obeys the density wave theory
(solid lines). If the motions of the Cepheids in the solar neighborhood
are in accordance with those expected in the density wave theory,
we expect the similar linear relation between $\phi$ and $\chi$.
However, such a linear relation is not apparent for the Cepheids, which may
imply that these objects in the region we investigate do not obey
the density wave theory.

To be more quantitative, we analyze the motions of the Cepheids by using
the sum of squares of the deviations for the observed $\phi$'s from
the theoretically expected ones.
We define this deviation, $\Delta^2$, as follows:
\begin{equation}
\Delta ^2 \equiv \sum _i \frac{\delta \phi_i^2}{\sigma_i^2} \ ,
\label{3}
\end{equation}
where $\delta \phi_i$ is defined as,
\begin{equation}
\delta \phi_i \equiv 
min( |\phi_i-\phi_{DW}(\chi_i)|,2\pi-|\phi_i-\phi_{DW}(\chi_i)| ) \ ,
\label{4}
\end{equation}
where $\sigma_i$ is the error in the phase $\phi_i$ of the star $i$,
and $\phi_{DW}(\chi_i)$ is the phase when the motion of the star $i$
obeys the density wave theory. Here we note that the error $\sigma_i$ includes
both the observational error and the velocity dispersion of the Cepheids.
We adopt 13 km s$^{-1}$ for the velocity dispersion.
The quantity $\delta \phi_i$ denotes the discrepancy between $\phi_i$ and 
$\phi_{DW}(\chi_i)$, for which we take the smaller value of 
$|\phi_i-\phi_{DW}(\chi_i)|$ and $2\pi-|\phi_i-\phi_{DW}(\chi_i)|$,
because $\phi_i$ has a period of $2\pi$.
The phase $\phi_{DW}(\chi_i)$ is determined by minimizing the deviation
$\Delta^2$ for each value of the arm number $m$.

If the motions of stars are in accordance with those expected in the density
wave theory, the expected value of the deviation $\Delta^2$ is about the same
as the number of the sample stars (78 in this work), because the distribution
of $\delta \phi_i^2$ is normalized by its standard dispersion $\sigma_i^2$.
Thus, in this case, we will obtain a minimum of $\Delta^2$ with
$\Delta_{min}^2 \la 80$ at a specific arm number $m = m_g$, whereas
at other $m$'s, the value of $\Delta^2$ will be systematically larger than 80;
the larger ratio of $\Delta^2 / \Delta_{min}^2$ implies more likely the
observed stellar motions match the density wave theory.
On the other hand, if $\Delta^2$ is always larger than 80 at all $m$'s,
we may conclude that the stellar motions are totally inconsistent with 
those expected in the theory. 

We plot the deviation $\Delta^2$ as a function of $m$ in Figure 4.
Solid line shows the result for our Cepheid sample.
It is apparent that $\Delta^2$ is around $250$ for all values of the parameter
$m$, without showing a noticeable minimum with $\Delta^2 \sim 80$. 

In order to examine the significance of this result, we investigate three
hypothetical models for comparison. As the first model, we randomly select
$\phi$, independently of $\chi$, so that there is no correlation between
$\phi$ and $\chi$. We then assign the same errors $\sigma_i$ as those for the
Cepheids to each $\phi$. Thick dashed line in Figure 4 shows $\Delta^2$ vs. $m$
derived from this ``random model''. It follows that the properties of
$\Delta^2$ as a function of $m$ are basically the same as for the Cepheids,
thereby implying that the phase $\phi$ of the Cepheids is random in the solar
neighborhood.
The second model, as shown by thin dashed line, follows the density wave
theory with $m=4$. We assign the same errors $\sigma_i$ as those for the
Cepheids. The value of $\Delta^2$ is around $250$, except for the case of
$m=4$ at the minimum of $\Delta^2$ ($\Delta_{min}^2 \sim 80$). This behavior
of $\Delta^2$ is in sharp contrast to the case of the Cepheids.
The last model (dotted line) is similar to the second one, except for the
assignment of larger errors $\sigma_i$: 1.4 times as much as those of
the Cepheids, in order to see the effect of $\sigma_i$ on the result.
In this case, the value $\Delta^2$ is smaller than other cases:
$\Delta \sim 150$ except for the case of $m=4$ at the minimum of $\Delta^2$
($\Delta_{min}^2 \sim 80$). Thus, even if the errors $\sigma_i$ are large,
which gives rise to small $\Delta^2/\Delta^2_{min}$,
$\Delta^2$ has a noticeable minimum at a specific value of $m$, 
in sharp contrast to the case of the Cepheids.

Therefore, these results suggest us that the motions of the Cepheids in the
solar neighborhood are inconsistent with those expected in the
density wave theory, whereas the random model, in which the phases
$\phi$ and $\chi$ are randomly selected, reproduces the observation in a
reasonable manner.

\section{DISCUSSION AND CONCLUSIONS}
We have presented a method, based on the analysis of local kinematics of disk
stars, to clarify whether the motions of the stars, which make up the spiral
arms, follow those expected in the density wave theory. The method utilizes
the comparison with the expected linear relation between the
``position phases'' of the stars $\chi$ and those of their epicyclic motions
$\phi$, as given in equation (\ref{2}). The application of the method to
the 78 Galactic Cepheids within 4 kpc from the Sun, for which accurate proper
motions are available from {\it Hipparcos}, has revealed that the relation
between $\chi$ and $\phi$ for the Cepheids does not show the expected linear
relation. Based on the quantitative analysis using the deviation $\Delta^2$,
we conclude that the observed motions of the Cepheids are well reproduced
by the random model having no correlation between $\chi$ and $\phi$.

There are a couple of possibilities to explain the current results, even if
the spiral arms follow the density wave theory. First, the spiral structure
around the Sun may not be simple as given in equation (\ref{1}). In fact,
many of our sample Cepheids belong to the local spiral structure called the
Orion arm, for which the definite conclusion on its spatial structure is
yet to be reached. It is frequently expressed as ``Orion spur'', having
a rather irregular pattern compared to other large-scale arms, Sagittarius
and Perseus arm (see, e.g., Gilmore, King, \& van der Kruit 1989). The
existence of the Orion arm may give disturbances on the density wave motions
of stars induced by these large-scale arms.
Second, the Cepheids we have adopted here may still convey systematic
velocities of dens gas clouds from which these stars were formed, in the form
of the streaming motions. If there still exist some individual streaming
motions among the sample stars, such motions may violate the ideal linear
relation between $\chi$ and $\phi$ expected for the density wave motions.
Third, the Cepheids we have adopted here may have already experienced some
scattering by dens gas clouds, thus having large velocity dispersions
(Spitzer \& Schwarzschild 1953). However, our experiment in \S 4 
(dotted line in Figure 4) implies that
the effect of velocity dispersions of our sample on the result appears to
be minor.

In order to settle the last issue described above more clearly, we have
repeated our analysis using younger populations with smaller velocity
dispersions than the Cepheids. As such young stars, we have adopted
the O-B5 stars,
although due to their fainter luminosities than Cepheids, the sample with 
available proper motions is confined to the narrower region near the Sun.
These sample stars are taken from 
the NASA SKY2000 Master Star Catalog Ver.~2 (Sande et al. 1998)
which provides almost
300,000 stars brighter than 8 mag. The catalog contains many basic
quantities, such as MK classification, luminosity class, apparent magnitude,
color, radial velocity, and so on. We have then calibrated distances using
{\it Hipparcos} parallaxes or spectroscopic distances using the program
kindly supplied by Drs. M.~S\^oma and M.~Yoshizawa, and also obtained
accurate proper motions by the cross-identification with the {\it Hipparcos}
and ACT Reference Catalogs (Urban, Corbin, \& Wycoff 1998).
After removing binaries and multiples, we have selected 773 O-B5 stars for
which full three-dimensional velocities are available. Then, the application
of the method we have developed here has revealed that the deviation
$\Delta^2$ as a function of $m$ remains essentially constant of the order of
3000, without showing any noticeable minimum at a
specific value of $m$. Thus, even the motions of such young populations with
small velocity dispersions are in contradiction with the density wave theory.
We note here that most of these O-B5 stars are located within $\sim 1$ kpc
from the Sun, so the effect of the local irregular spiral on the result 
cannot be negligible.

More definite conclusions on the issue we have addressed here require the
assembly and analysis of much larger numbers of stars with accurate distances
and proper motions, so that the statistical fluctuation in the result can
be significantly reduced. Also, it is necessary to assemble the data of
more remote stars over a large fraction of the disk, thereby diminishing
effects of local irregular spiral structures on the kinematic analysis.
Indeed, next-generation satellites such as {\it FAME} and {\it GAIA} will
provide very precise astrometric data for huge numbers of the
Galactic stars, and will thus offer us an opportunity to assess detailed
motions of disk stars in conjunction with the density wave theory.

\acknowledgments  
We are grateful to M. S\^oma and M. Yoshizawa for providing us the program
for calculating spectroscopic distances. T.Y. would like to thank K. Okoshi
and  S. Bouquillon for useful discussion. This work was supported in part by
Research Fellowships of the Japan Society for the Promotion of Science 
for Young Scientists (No.00074).

\clearpage

\begin{figure}
\plotone{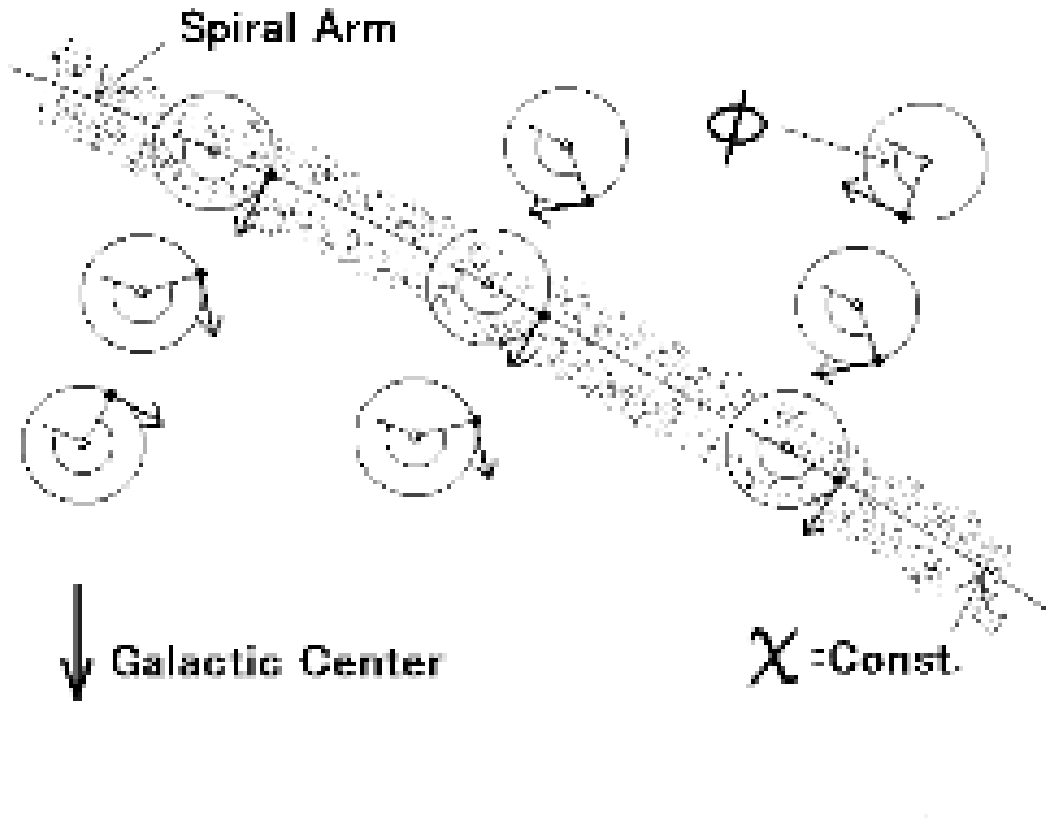}
\caption{Schematic diagram showing the definition of phases, $\chi$ and $\phi$,
in the Galactic Plane. Each vector corresponds to a velocity vector of
each star in epicyclic motion.}
\end{figure}

\begin{figure}
\plotone{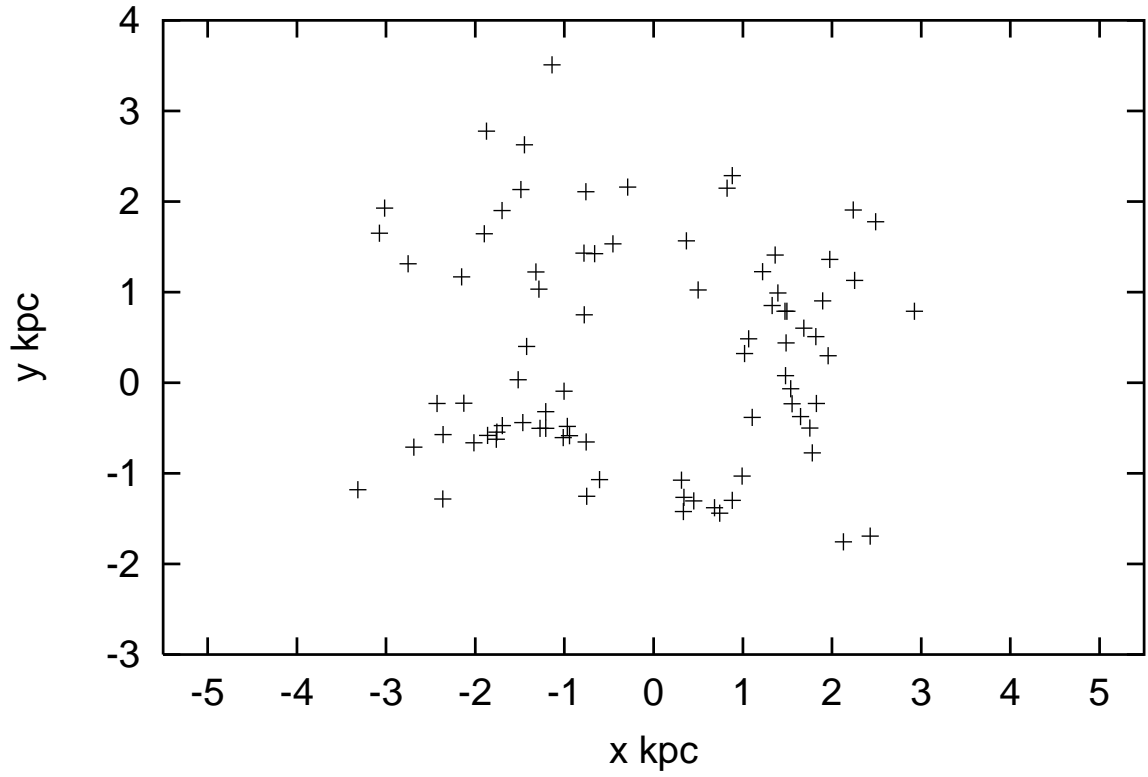}
\caption{Positions of the Cepheids in the Galactic plane. The position 
of the Sun is the origin of the coordinate axes.
The Galactic Center is located in the negative direction of 
the $y$-axis with $x=0$.}
\end{figure}

\begin{figure}
\plotone{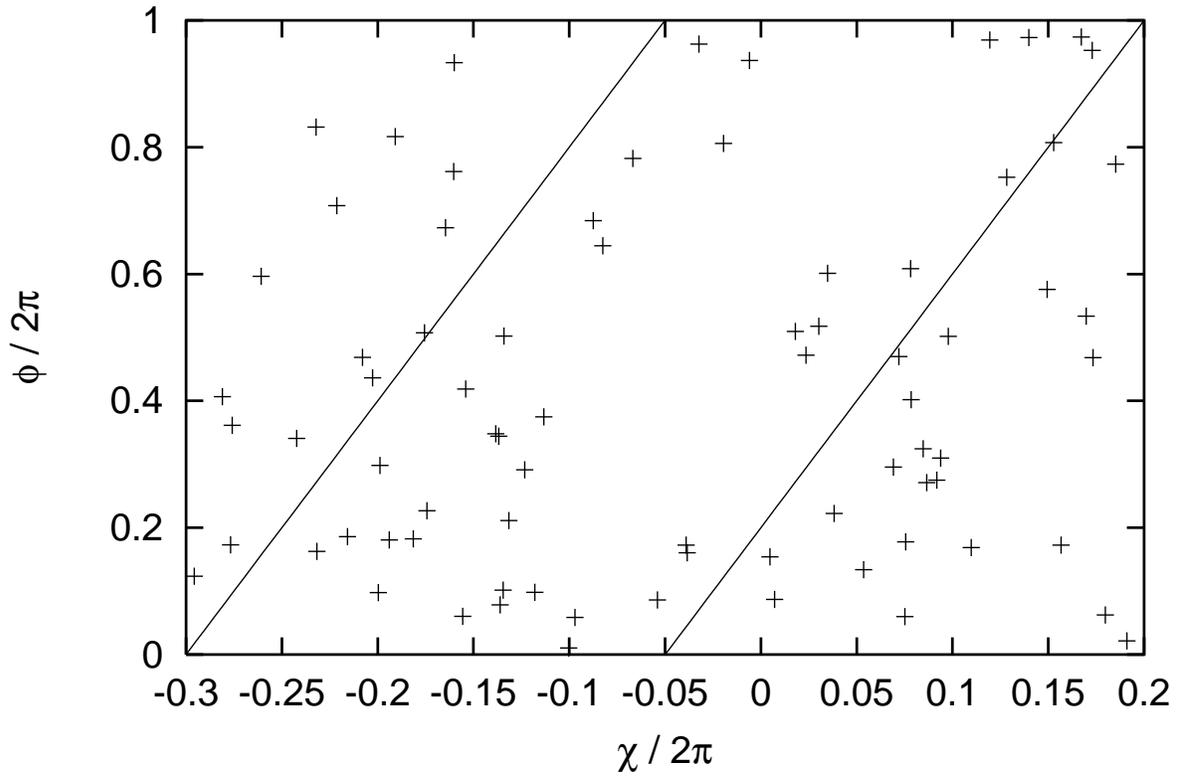}
\caption{ 
The phase $\phi /2\pi$ is plotted against $\chi/2\pi$ for our sample of
Cepheids. The solid lines show the relation expected for a 4-armed galaxy
that obeys the density wave theory.}
\end{figure}

\begin{figure}
\plotone{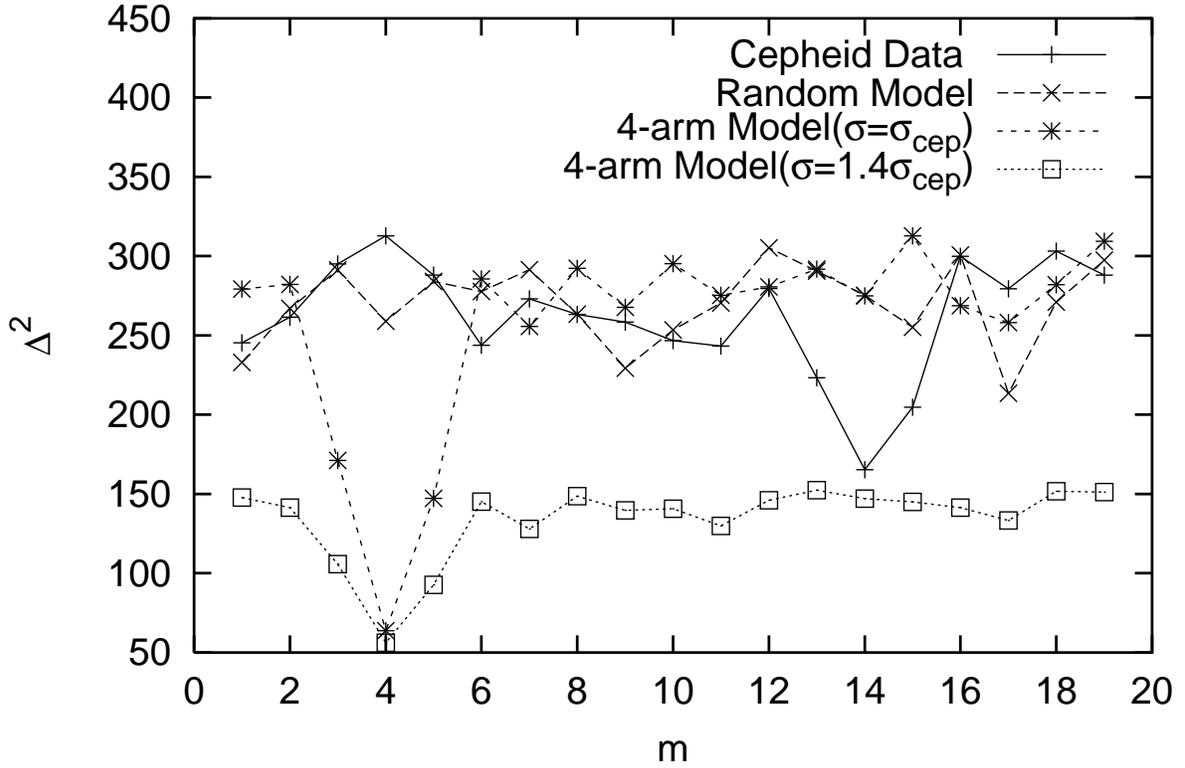}
\caption{The sum of square of the deviations, $\Delta^2$, as a function of
arm-number $m$. The solid, dashed, thin dashed, and dotted lines denote
Cepheid data, random model, 4-arm model ($\sigma=\sigma_{cep}$),
and 4-arm model ($\sigma=1.4\sigma_{cep}$), respectively. Here $\sigma_{cep}$
stands for the velocity error in the original Cepheid sample.}
\end{figure}

\end{document}